\documentclass[prd,showpacs,twocolumn,nofootinbib]{revtex4-1}
\usepackage{amsfonts,amsmath,amssymb,amsthm}
\usepackage[bookmarks, bookmarksopen, bookmarksnumbered]{hyperref}
\usepackage{graphicx}

\newcommand{\de}{\mathrm{d}}
\newcommand{\dd}[3]{\frac{ \de^{ #1 } #2 }{ \de #3^{ #1 } }} 

\begin{document}
\title{Modelling general-relativistic disk in OJ 287} 
\author{Wojciech Dyba}
\author{Patryk Mach}
\author{Edward Malec}

\affiliation{Instytut Fizyki im.\ Mariana Smoluchowskiego, Uniwersytet Jagiello\'{n}ski,
{\L}ojasiewicza 11, 30-348 Krak\'{o}w, Poland}

\begin{abstract}
We model self-gravitating disks in Keplerian motion around the primary black hole, in the binary black hole system OJ 287 with a torus, employing a consistently general-relativistic approach. They satisfy geometric and/or mass density requirements found by Sillanp\"{a}\"{a}, Valtonen, Lehto and their coworkers. It is plausible that essential observational features of OJ 287 can be obtained via the general-relativistic description of the Bondi-Hoyle-Lyttleton transits through these tori, within the framework of radiation hydrodynamics. 
\end{abstract}

\maketitle

\section{Introduction}
OJ 287 is an active galactic nucleus (AGN) in the Cancer constellation. It has been identified in the sky photographs dating from 1880's. The light curve of OJ 287 has a quasi-periodic variability --- there appears a pair of flares at every 12 years with intensity modulated in a 60 year period.

A.\ Sillanp\"{a}\"{a} et al.\ \cite{BBHmodel1988} interpreted this object as a binary black hole consisting of a supermassive central black hole and a much lighter companion, with an accretion disk surrounding the central object. This model has been further developed by Lehto and Valtonen \cite{LehtoValtonen1996}. The accretion disk has been modelled as a version of the classical Shakura-Sunyaev thin disk model \cite{Shakura_Sunyaev} due to Sakimoto and Coroniti \cite{Sakimoto}. The appearance of flares is explained in this model as a consequence of piercing the accretion disk around the primary most massive black hole by the smaller one. Bondi-Hoyle type arguments imply the production of radiation during the transition process \cite{HL,BH,Edgar}. The resultant radiation flare is thermal and it is followed by nonthermal radiation flares generated by subsequent jet perturbances \cite{model2018}. The overall luminosity increases significantly during these thermal outbursts. The whole system is compact. In particular, the accretion disk extends from several Schwarzschild radii $R_\mathrm{S} \equiv 2GM/c^2$ to a few dozens of $R_\mathrm{S}$. The orbit of the smaller black hole has a similar size. The vertical width of the disk is estimated to be of the order of $R_\mathrm{S}$. This model is supported by several observational arguments and has been used for tests of General Relativity in several post-Newtonian orders \cite{Valtonen2008testGR,Pihajoki}.

The model of \cite{BBHmodel1988} assumes implicitly that the analysis of the 3-object system --- two black holes and a disk --- can be reduced to three independent parts: the 2-body interaction between the black holes, the interaction between a torus and the primary black hole, and the quasi-periodic encounters of the disk and the secondary black hole. This assumption might be valid, if the disk mass $ M_\mathrm{d}$ is significantly smaller than the mass of the secondary black hole.

The purpose of this paper is to determine the general-relativistic equilibria of polytropic tori coupled to black holes. Strictly saying, we consider the primary black hole --- torus system, thus neglecting (as in the former investigations, cf.\ \cite{BBHmodel1988,BBHmodel1988,LehtoValtonen1996,P2013,Pihajoki,model2018}) the influence of the smaller black hole onto the structure of a torus. In the next section we shall describe the needed formalism. Section \ref{numres} presents numerical results concerning tori;  a few of them are particularly interesting, because tori masses are much smaller than the mass of the secondary black hole. That suggests consistency of the approach proposed in \cite{BBHmodel1988, LehtoValtonen1996}. The potential significance of obtained solutions is discussed in Sec.\ \ref{discussion}. We note, in particular, that Zanotti et al.\ \cite{Rezzolla} have shown that the Bondi-Hoyle-Lyttleton accretion of gas onto the secondary black hole can be \textit{effectively} described within the framework of the general-relativistic radiation hydrodynamics. Their analysis is quite general, it is not meant to apply to any specific system, but it yields a luminosity estimate that essentially agrees with observations of OJ 287. The last section is devoted to a brief summary.

\section{On equations, numerical methods and physical quantities}
 
The content of this section is based on \cite{pracanum}, which in turn is just a (slight) reformulation of Shibata \cite{Shibata}. The axially symmetric rotating spacetime with matter is described by the line element 
\begin{eqnarray}
\de s^2 &=& -\alpha^2 \de t^2 + r^2 \sin^2 \theta \, \psi^4 (\de \varphi + \beta \de t )^2 +\nonumber\\
&& \psi^4 e^{2q} (\de r^2 + r^2 \de \theta^2).
\label{1}
\end{eqnarray}
Here metric functions $\alpha$, $q$, $\beta$ and the conformal factor $\psi $ depend only on $r$ and $\theta$. We adopt standard general-relativistic units with $G=c=1$. We assume a polytropic gas with the stress-energy tensor
\begin{equation}
T^{\alpha \beta} = \rho h u^{\alpha} u^{\beta} + p g^{\alpha \beta}.
\end{equation}
The equation of state relates the pressure $p$ and the  baryonic rest mass density $\rho$, $p(\rho) = K \rho^{\gamma}$. The specific enthalpy can be put in the form
\begin{equation}
h(\rho) = 1+ \frac{\gamma p}{(\gamma-1)\rho}.
\end{equation}

The 4-velocity of particles of the fluid is given by $(u^{\alpha})=  (u^t, 0, 0, u^{\varphi})$. The coordinate angular velocity of rotating particles is equal to
\begin{equation}
\Omega = \frac{u^{\varphi}}{u^{t}}.
\end{equation}
We shall define the angular momentum per unit inertial mass $\rho h$, 
\begin{equation}
j \equiv u_{\varphi} u^{t}.
\end{equation}
It is well known, since Bardeen \cite{Bardeen_1970} that --- under the assumptions of stationarity and axisymmetry  --- there exists an integrability condition $j=j(\Omega )$. In such a case one has an integro-algebraic Bernoulli-type equation
\begin{equation}
\int h u_{\varphi} \de \Omega + \frac{h}{u^{t}} = C_{1}
\end{equation}
or
\begin{equation}
\int u^{t} u_{\varphi} \de \Omega + \ln \left( \frac{h}{u^{t}} \right)= C_{2},
\label{IntArgBerEq}
\end{equation}
where $C_1$ and $C_2$ are constants. The formulations employing one or the other of the two equations are  equivalent. We shall use Eq.\ (\ref{IntArgBerEq}) in our calculations.
 
The rotation laws of fluid tori constituting general-relativistic versions of Newtonian monomial  rotations  have been discovered quite recently \cite{pracanum,MM2015,KKMMOP}. In particular, the general-relativistic version of the Keplerian rotation around a spin-less central black hole is given by
\begin{equation}
j(\Omega) \equiv \frac{{w}^{4/3} \Omega^{-1/3}}{1-3 {w}^{4/3} \Omega^{2/3}}
\label{nspin}.
\end{equation}
Fluids rotating around  spinning black holes obey  the general-relativistic Keplerian rotation law
\begin{equation}
j(\Omega) = - \frac{1}{2 } \dd{}{}{\Omega} \ln \left\{ 1-  \left[ \tilde a^2 \Omega^2 + 3 w^{\frac{4}{3}} \Omega^{\frac{2}{3}}(1-\tilde a \Omega)^{\frac{4}{3}}\right]\right\}.
\label{rotation}
\end{equation}
Here $\tilde a$ is a kind of a spin  parameter and $w^2$ is a mass. It coincides, for massless tori, with the spin parameter of the Kerr black hole. Obviously (\ref{rotation}) coincides with (\ref{nspin}) for $\tilde a=0$. In the rest of this paper we shall use the rotation curve (\ref{rotation}).

Let us point that the ``usual'' form of the angular velocity --- that is, angular velocity as a function of spatial coordinates --- can be obtained, provided that the metric is known. It can be recovered from the equation 
\begin{equation}
j(\Omega)= \frac{V^2}{(\Omega + \beta)(1-V^2)},
\end{equation}
where 
\begin{equation}
V^2 = r^2 \sin^2 \theta (\Omega+\beta)^2 \frac{\psi^4}{\alpha^2}.
\end{equation}

In the Newtonian limit $V^2/c^2 \rightarrow 0$, assuming a test disk approximation and restoring the SI units, one would find the Keplerian angular velocity $\Omega =\frac{\sqrt{Gm}}{(\sin \theta r)^{3/2}}$, where $m$ would be the central mass.

We need to define quasilocal characteristics of the central black holes. This is done in a standard way. The angular momentum is given by the Komar integral over the boundary of the black hole,
\begin{equation}
J_{\mathrm{H}} = \frac{1}{4} \int^{\pi/2}_{0} \frac{r^4 \psi^6}{\alpha} \partial_{r} \beta \sin^3 \theta \de \theta .
\end{equation}
   
The mass is defined in two steps. First, one defines the irreducible mass of black hole, as being related to the area $A$ of the event horizon of the black hole: 
\[M_{\mathrm{irr}}= \sqrt{\frac{A_{\mathrm{H}}}{16 \pi}}. \]
The mass of the black hole is then given by
\begin{equation}
M  = M_{\mathrm{irr}} \sqrt{1+\frac{J^2_{\mathrm{H}}}{4 M^4_{\mathrm{irr}}}}.
\end{equation}
Thus the mass $M$ is expressed in terms of two quasi-local characteristics of the event (apparent) horizon --- its area and its angular momentum.
 
The spin parameter of the black hole is given by  
\[ a =\frac{J_{\mathrm{H}}}{ M }. \]
It is equal to the bare spin parameter $\tilde a$ in the rotation law (\ref{rotation}) in the case of massless tori, when the geometry coincides with the Kerr geometry.

The asymptotic characteristics --- the mass $M_\mathrm{ADM}$ and the total angular momentum $J_\mathrm{ADM}$ --- are defined in the standard way, as certain boundary expressions at spatial infinity \cite{Shibata}.
 
There exists a possibility to define a quasilocal mass of tori, but we choose the simpler option \cite{pracanum}
\begin{equation}
m_\mathrm{T} \equiv M_{\mathrm{ADM}} - M;
\end{equation}
the mass of a torus is just a difference between the asymptotic mass of the spacetime and the quasilocal mass of the black hole.

We shall not write here the relevant Einstein equations; they can be found in \cite{Shibata} and \cite{pracanum}. For a detailed description of our numerical method see \cite{pracanum}; this method deviates in a few places from the approach of Shibata \cite{Shibata}.

A few comments are in order. From the mathematical point of view finding a toroid around a black hole is a free boundary problem. One cannot specify the shape of the toroid; it emerges as a part of the solution. It is customary, in numerical calculations, to specify only the location of the inner and the outer disk boundaries, at the symmetry plane $\theta =\pi /2$. The other datum is the maximal mass density inside the torus, or (equivalently) its mass. One should specify also characteristics of the central black hole: its bare mass parameter $m$, and the angular momentum $J_\mathrm{H}=\tilde a  m$. We shall measure $m$ in units of the mass $\mathrm{M}$ of the central black hole --- thus $m=1$, and the spin $J_\mathrm{H}=\tilde a  m=\tilde a$. These data are complemented by the equation of state of the fluid and the rotation law (\ref{rotation}). For details see \cite{pracanum}.

\section{Numerical results}
\label{numres}

We base our numerical calculations on data given in publications of Valtonen and collaborators \cite{LehtoValtonen1996,P2013,model2018}. In what follows $\mathrm{M}$ is the mass of the central black hole, $\mathrm{M}=1.8348\times 10^{10} \, \mathrm{M}_\odot$. The secondary black hole has a mass $M_\mathrm{sbh}=1.5\times 10^{8} \, \mathrm{M}_\odot$. We put the spin $J_\mathrm{H} = 0.37 m = 0.37$ of the primary black hole, and the values $w=1$, $\tilde a = 0.37$  in the rotation law (\ref{rotation}).

Valtonen et al.\ have assumed that innermost and outermost boundaries of their disks are located --- on the symmetry plane $\theta =\pi/2$ --- around circumferential radii $R_\mathrm{C1}=20 \, \mathrm{M}$ (or 3600 astronomical units --- AU in what follows) and $R_\mathrm{C2}=120 \, \mathrm{M}$ (or 20600 AU), respectively. They take, as additional parameters, a maximal density $\rho_\mathrm{m}$ and a height $h$ of a disk. The maximal mass density is comprised within the interval $(1.33 \times 10^{-7}, 2.5 \times 10^{-7}) \, \mathrm{kg}/\mathrm{m^3}$, while the half-thickness of disks is within the range $(1.2, 1.4) \, \mathrm{M}$ (or 220--260 AU).

Disks that have been modelled in \cite{LehtoValtonen1996,P2013,model2018} constitute versions, constructed by Sakimoto and Coroniti \cite{Sakimoto}, of the Newtonian thin and massless disks of Shakura and Sunyaev \cite{Shakura_Sunyaev}, rotating with the Keplerian angular velocity $\Omega \propto 1/r^{3/2}$. They are applied within the general-relativistic context; we find this procedure as inherently inconsistent. The purpose of this Section is to construct general-relativistic toroids that are as close as possible to disks postulated by Valtonen et al.
 
We will get general-relativistic disks for two polytropic equations of state, $p=K\rho^{5/3}$ or $p=K\rho^{4/3}$. The equation of state $p=K\rho^{4/3}$ can be regarded as an effective equation of state that takes into account both the radiation and baryonic matter; see a discussion in \cite{Mihalas,ZR}.  

It is necessary to say, that in the modelling of polytropic tori, one cannot impose simultaneously the height and the maximal mass density of tori; one can use one of them, but almost never two, since otherwise solutions do not exist. Therefore we discuss separately two cases, one with the datum $h$ and the other with $\rho_\mathrm{m}$. In one class of solutions we fix the geometric extent by prescribing circumferential radii $R_\mathrm{C1}$ and $R_\mathrm{C2}$ in the symmetry plane $\theta =\pi/2$  and the maximal mass density $\rho_\mathrm{m}$. The numerical method of \cite{pracanum} is specifically elaborated for this set of boundary data. In the other set of solutions we choose the circumferential radii $R_\mathrm{C1}$ and $R_\mathrm{C2}$ in the symmetry plane $\theta =\pi/2$, and we specify the coordinate maximal height $h$. The desired value of the parameter $h$ is obtained by the method of trial and error, by a suitable choice of the maximal mass density $\rho_\mathrm{m}$. \textit{We should acknowledge, that the proper geometric quantity in this height-related scheme would be the maximal geodesic height of the disk, instead of the coordinate one. This in principle could be implemented, but at a high computational cost. Fortunately, it appears that the maximal geodesic height is close to the coordinate one in the forthcoming numerical examples, and for that reason we used a simplified numerical setting with the coordinate height $h$.} In all cases we put these quantities within the ranges that are regarded as being of interest in \cite{LehtoValtonen1996,P2013} and \cite{model2018}.

We repeat these calculations for yet another set of boundary data. While keeping values of $R_\mathrm{C2}$ and $\rho_\mathrm{m}$ (or $h$), we minimize values of the circumferential radii $R_\mathrm{C1}$ of innermost disk's boundaries. This is done empirically, through the method of trial and guess in numerical calculations.    

\subsection{Toroids of proper height}
\label{numres1}

The exemplary solutions described in what follows have geometric dimensions as determined in \cite{model2018}; in particular, their maximal height $h$ is close to anticipated values $h \in  (150, 250) \, \mathrm{AU}$. We would like to point that $h/(R_\mathrm{C1} - R_\mathrm{C2})\approx 0.01$; the construction of such thin disks is a formidable numerical challenge. Typically one calculation required about 1000--2000 CPU hours and 0.128 terabyte of RAM. In all cases disks masses appeared to be much smaller than that of the secondary black hole. 

\subsubsection{The equation of state  $p=K\rho^{5/3}$}
\label{numres11}

\begin{figure}[ht]
\includegraphics[width=1\columnwidth]{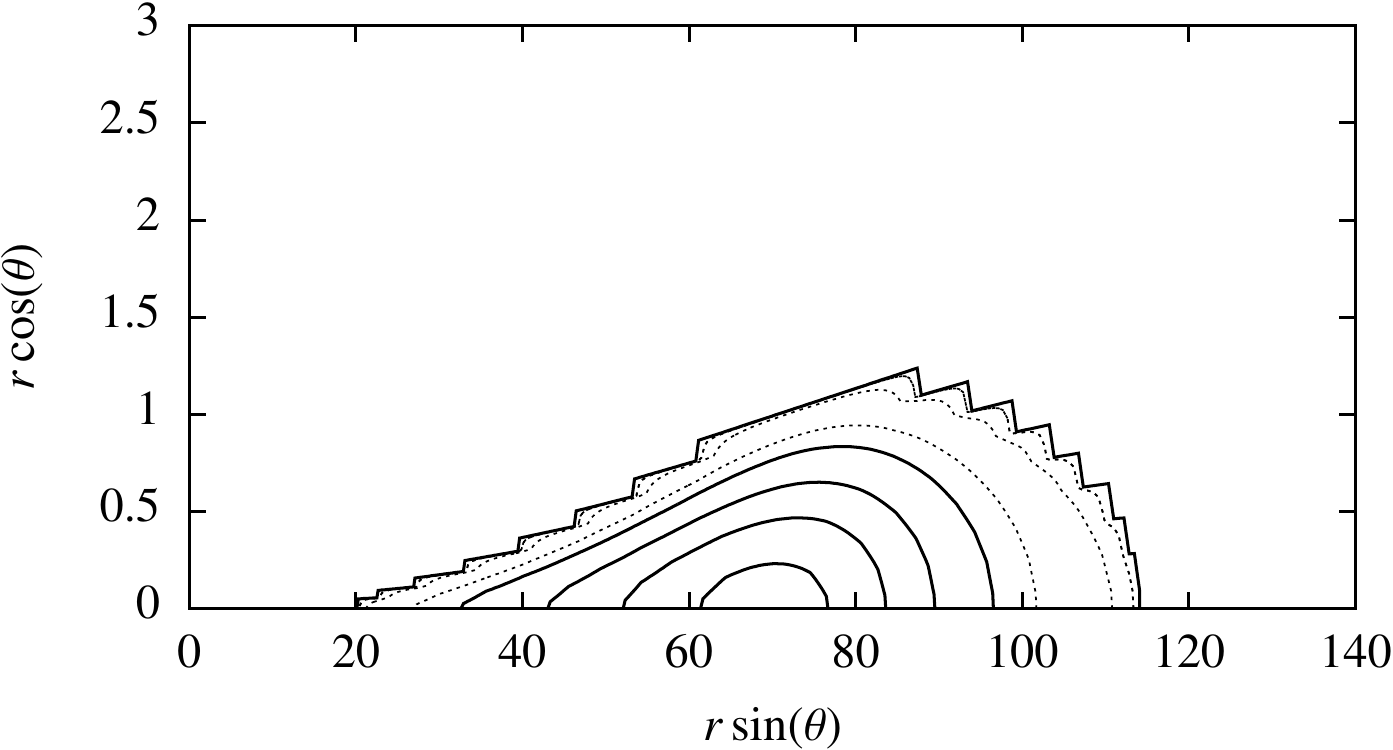}
\caption{\label{fig1} Here  $R_\mathrm{C1}=20.96 \, \mathrm{M}$, $R_\mathrm{C2}=115.10 \, \mathrm{M}$ and $\rho_\mathrm{m} =8.8\times 10^{-9}/\mathrm{M}^2$. Solid density isolines correspond to $\rho = (0, 2, 4, 6, 8) \times 10^{-9}/\mathrm{M}^2$ (linear scale). Dotted density isolines correspond to $\rho = (10^{-11}, 10^{-10}, 10^{-9})/\mathrm{M}^2$ (logarithmic scale).}
\end{figure}  

In Figure \ref{fig1} we assume the circumferential radius of the innermost disk boundary $R_\mathrm{C1}=20.96 \, \mathrm{M}$ (or 3795 AU) and the circumferential radius of the outermost  boundary $R_\mathrm{C2}=115.10 \, \mathrm{M}$ (20840 AU). In standard SI units we have $R_\mathrm{C1}=5.68\times 10^{14} \, \mathrm{m}$ and $R_\mathrm{C2}=3.12\times 10^{15} \, \mathrm{m}$. The mass density reads $\rho_\mathrm{m} =8.8\times 10^{-9}/\mathrm{M}^2$ ($\rho_\mathrm{m} = 1.62 \times 10^{-8} \, \mathrm{kg/m^3}$); this is about one tenth of the minimal mass density anticipated in \cite{LehtoValtonen1996} and \cite{model2018}.

The mass of the toroid is equal to $1.85\times 10^{-4} \, \mathrm{M} \approx 3.39\times 10^{6} \, \mathrm{M}_\odot$; it is small in comparison to the mass of the smaller black hole companion. Its maximal height $h=1.09 \, \mathrm{M}$ (or $197 \, \mathrm{AU} \approx 2.95 \times 10^{13} \, \mathrm{m}$) is close to the height expected in \cite{LehtoValtonen1996} and \cite{P2013, Pihajoki, model2018}.  

\begin{figure}[ht]
\includegraphics[width=1\columnwidth]{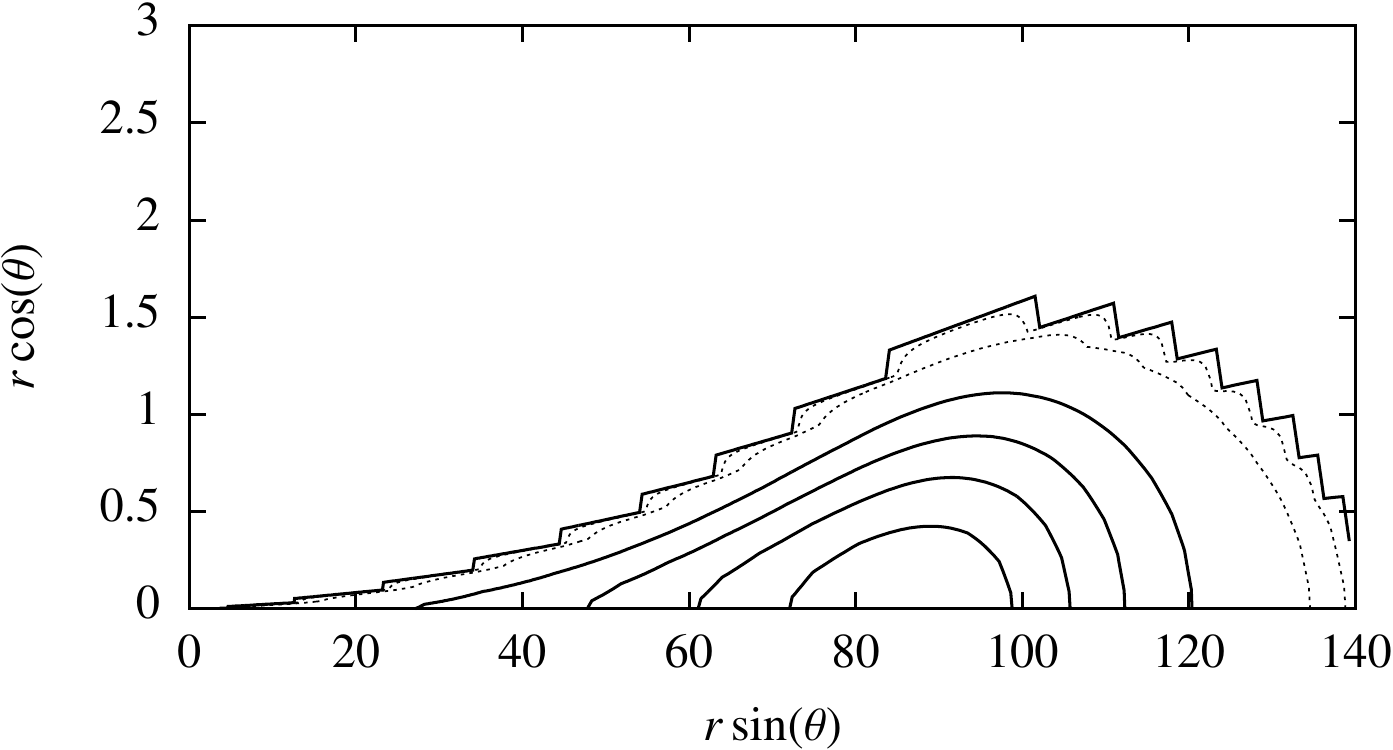}
\caption{\label{fig2} Here  $R_\mathrm{C1}=4.75 \, \mathrm{M}$, $R_\mathrm{C2}=141.03 \, \mathrm{M}$ and $\rho_\mathrm{m} =4.8\times 10^{-9}/\mathrm{M}^2$. Solid density isolines correspond to $\rho = (0, 1, 2, 3, 4) \times 10^{-9}/\mathrm{M}^2$ (linear scale). Dotted density isolines correspond to $\rho = (10^{-11}, 10^{-10})/\mathrm{M}^2$ (logarithmic scale).}
\end{figure}  
 
In the solution represented in Fig.\ \ref{fig2}, the circumferential radius of the innermost disk boundary $R_\mathrm{C1}=4.75 \, \mathrm{M}$ (or 860 AU) and the circumferential radius of the outermost boundary $R_\mathrm{C2}=141.03 \, \mathrm{M}$ (25535 AU). The adopted value $R_\mathrm{C1}$ is quite close to the circumferential radius of the innermost stable circular orbit; that is the main difference between Figs.\ \ref{fig2} and \ref{fig1}. In standard SI units we have $R_\mathrm{C1}=1.29\times 10^{14} \, \mathrm{m} $ and $R_\mathrm{C2}=3.82\times 10^{15} \, \mathrm{m} $. The mass density $\rho_\mathrm{m}$ reads $\rho_\mathrm{m} = 4.8\times 10^{-9} \, \mathrm{M}^2$  ($\rho_\mathrm{m} =8.8\times 10^{-9} \, \mathrm{kg/m^3}$); this is 1/15th of the minimal mass density adopted in \cite{LehtoValtonen1996} and \cite{model2018}.

The mass of the toroid is equal to $2.01\times 10^{-4} \, \mathrm{M}\approx 3.69 \times 10^{6} \, \mathrm{M}_\odot $; this is much less than the mass of the secondary black hole. Its maximal height $h=1.44 \, \mathrm{M}$ (or $261 \, \mathrm{AU} \approx 3.9\times 10^{13} \, \mathrm{m}$) is close to the estimate of \cite{LehtoValtonen1996} and \cite{model2018}.  
 
\subsubsection{The equation of state  $p=K\rho^{4/3}$}
\label{numres12}

\begin{figure}[ht]
\includegraphics[width=1\columnwidth]{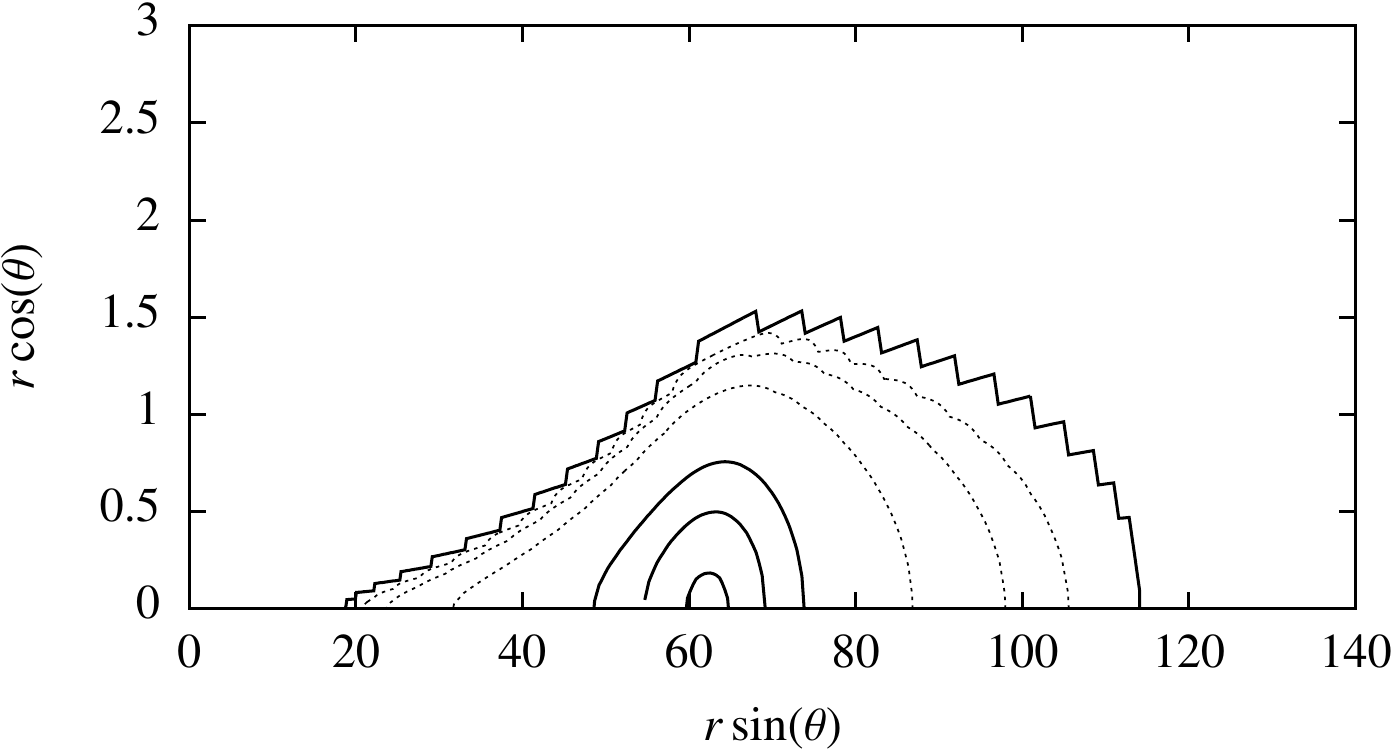}
\caption{\label{fig3} Here  $R_\mathrm{C1}=19.73 \, \mathrm{M}$, $R_\mathrm{C2}=115.09 \, \mathrm{M}$ and $\rho_\mathrm{m} =3.2\times 10^{-8}/\mathrm{M}^2$. Solid density isolines correspond to $\rho = (0, 1, 2, 3) \times 10^{-8}/\mathrm{M}^2$ (linear scale). Dotted density isolines correspond to $\rho = (10^{-11}, 10^{-10}, 10^{-9})/\mathrm{M}^2$ (logarithmic scale).}
\end{figure}  
   
The profile of the third solution is displayed in Fig.\ \ref{fig3}. The circumferential radii of the innermost and outermost disk boundaries are given by $R_\mathrm{C1}=19.73 \, \mathrm{M}$ (or 3570 AU) and $R_\mathrm{C2}=115.09 \, \mathrm{M}$ (20840 AU), respectively. In standard SI units we have $R_\mathrm{C1}=5.34\times 10^{14} \, \mathrm{m} $ and $R_\mathrm{C2}=3.12\times 10^{15} \, \mathrm{m} $. The mass density $\rho_\mathrm{m}$ reads $\rho_\mathrm{m} =6.8\times 10^{-9}/\mathrm{M}^2$ (or $1.25\times 10^{-8} \, \mathrm{kg/m^3}$); this is about one tenth of the minimal mass density assumed in \cite{LehtoValtonen1996} and \cite{model2018}.

The mass of the toroid is equal to $2.13\times 10^{-4} \, \mathrm{M}\approx 3.91\times 10^{6} \, \mathrm{M}_\odot$; its mass is small, as it should be, in comparison to the smaller BH companion. Its maximal height $h=1.06 \, \mathrm{M}$ (or $192 \, \mathrm{AU} \approx 2.87 \times 10^{13} \, \mathrm{m}$) is somewhat smaller than estimated in \cite{LehtoValtonen1996} and \cite{P2013, Pihajoki, model2018}.  

\subsection{Proper density toroids: $p=K\rho^{5/3}$ and $p=K\rho^{4/3}$}
\label{numres2}
 
In all forthcoming examples we assume $\rho_\mathrm{m} = 7.25\times 10^{-8}/\mathrm{M}^2$ ($\rho_\mathrm{m} =1.33\times 10^{-7} \, \mathrm{kg/m^3}$); this is the minimal mass density according to the discussion of \cite{LehtoValtonen1996} and \cite{model2018}. The horizontal sizes are close to data of \cite{model2018}, but the vertical size (the height) that results from numerical calculations, is much larger. In first and second cases the toroids are heavier than the secondary black hole.
 
\begin{figure}[ht]
\includegraphics[width=1\columnwidth]{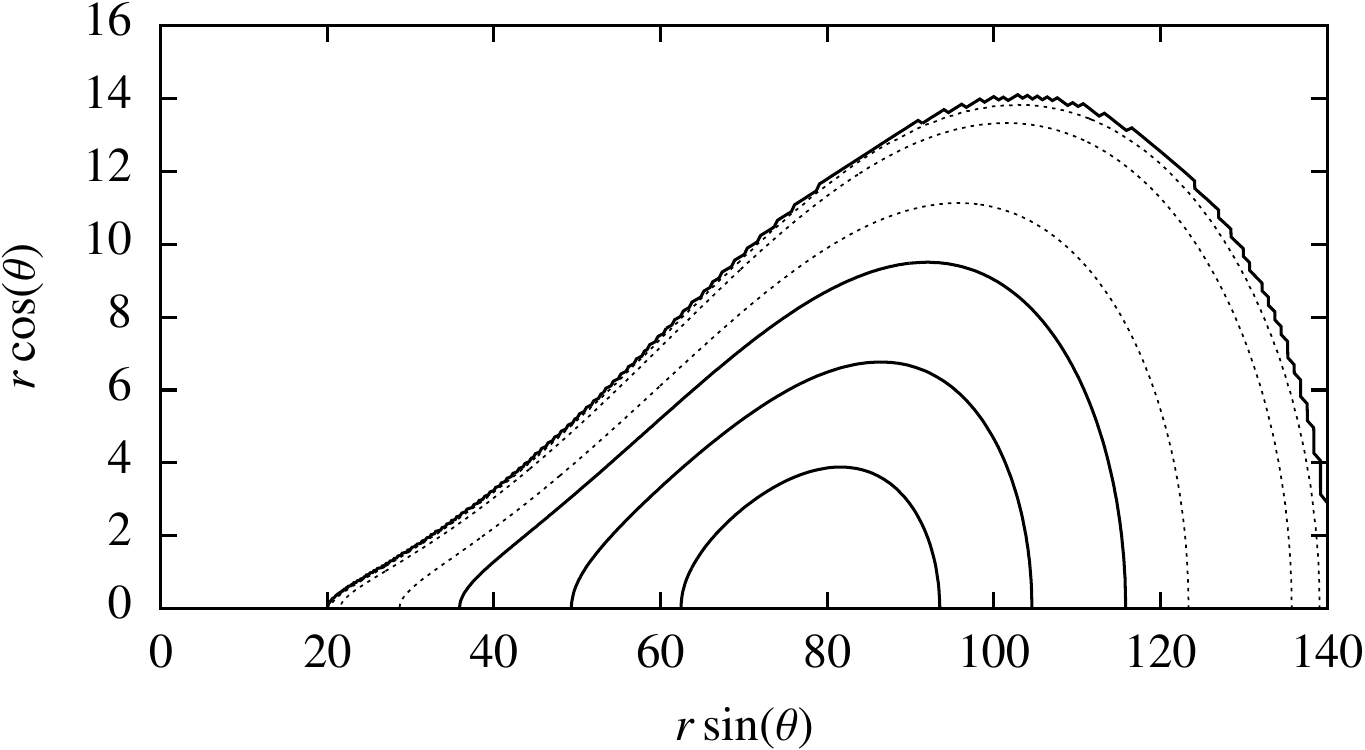}
\caption{\label{fig4} Here $R_\mathrm{C1}=21.08 \, \mathrm{M}$, $R_\mathrm{C2}=141.06 \, \mathrm{M}$.   The equation of state  $p=K\rho^{5/3}$. Solid density isolines correspond to $\rho = (0, 2, 4, 6) \times 10^{-8}/\mathrm{M}^2$ (linear scale). Dotted density isolines correspond to $\rho = (10^{-10}, 10^{-9}, 10^{-8})/\mathrm{M}^2$ (logarithmic scale).}
\end{figure}  
   
For Figure \ref{fig4} we assume the circumferential radius of the innermost disk boundary $R_\mathrm{C1}=21.08 \, \mathrm{M}$ (or 3816 AU) and the circumferential radius of the outermost boundary $R_\mathrm{C2}=141.06 \, \mathrm{M}$ (25540 AU). In standard SI units we have $R_\mathrm{C1}=5.71\times 10^{14} \, \mathrm{m} $ and $R_\mathrm{C2}=3.82\times 10^{15} \, \mathrm{m} $, respectively. The resulting mass $M_\mathrm{d}$ of the toroid is equal to $2.79\times 10^{-2} \, \mathrm{M}\approx 5.11\times 10^{8} \, \mathrm{M}_\odot$; that weight of more than a half of a billion solar masses is about the triple of the mass of the secondary black hole. The height of the torus $h=13.92 \, \mathrm{M}$ (or $2520 \, \mathrm{AU} \approx 3.77 \times 10^{14} \, \mathrm{m}$) exceeds tenfold expectations of \cite{LehtoValtonen1996} and \cite{P2013, Pihajoki, model2018}.  
   
\begin{figure}[ht]
\includegraphics[width=1\columnwidth]{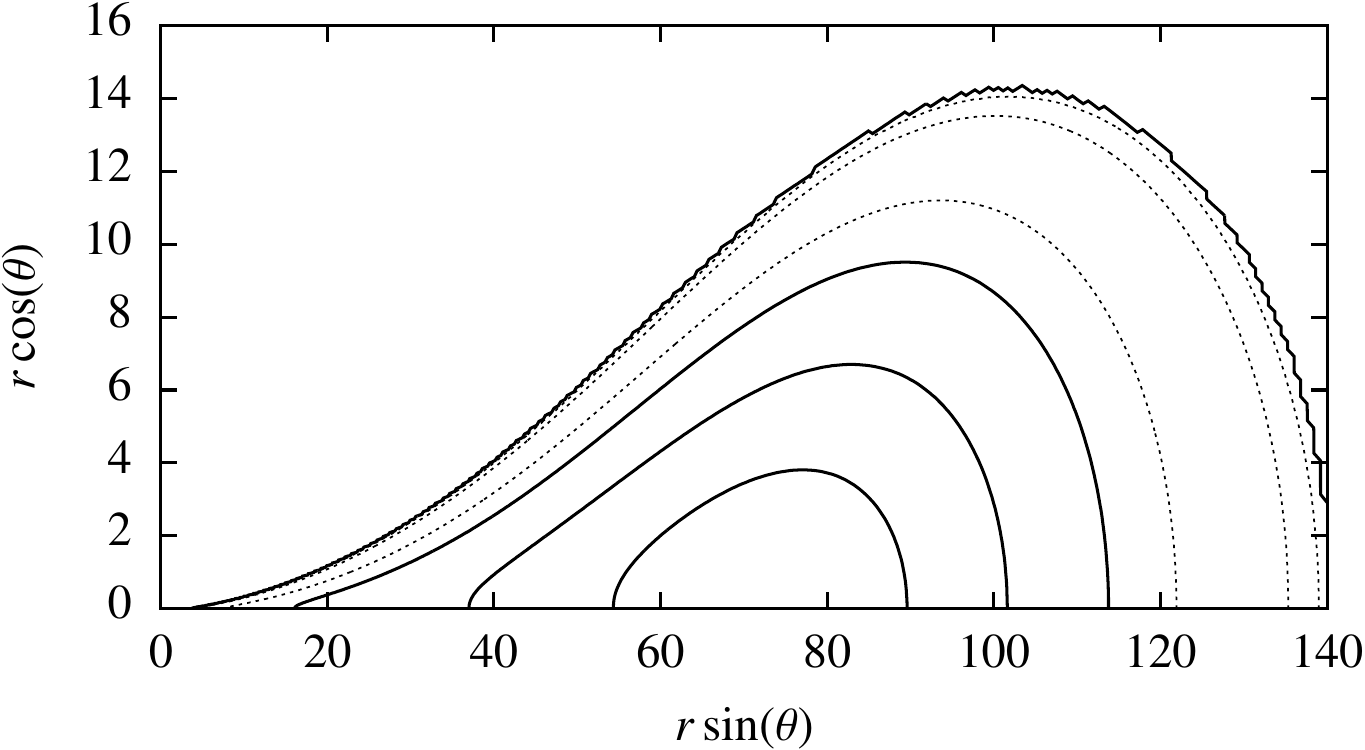}
\caption{\label{fig5} Here  $R_\mathrm{C1}=4.75 \, \mathrm{M}$, $R_\mathrm{C2}=141.06 \, \mathrm{M}$ and the equation of state  $p=K\rho^{5/3}$. Solid density isolines correspond to $\rho = (0, 2, 4, 6) \times 10^{-8}/\mathrm{M}^2$ (linear scale). Dotted density isolines correspond to $\rho = (10^{-10}, 10^{-9}, 10^{-8})/\mathrm{M}^2$ (logarithmic scale).}
\end{figure}

In Figure \ref{fig5} we find  the smallest possible value of the circumferential radius of innermost disk boundary, $R_\mathrm{C1}=4.75 \, \mathrm{M}$ (or 860 AU). The radius of the outermost disk boundary reads $R_\mathrm{C2}=141.06 \, \mathrm{M}$ (25536 AU). In standard SI units we have $R_\mathrm{C1}=1.30\times 10^{14} \, \mathrm{m} $ and $R_\mathrm{C2}=3.82\times 10^{15} \, \mathrm{m}$. The mass of the toroid is equal to $2.85\times 10^{-2} \, \mathrm{M} \approx 5.23 \times 10^{8} \, \mathrm{M}_\odot$, that is more than a half of  billion solar masses. Its height $h=14.18 \, \mathrm{M}$ (or c.\ 2567 AU) exceeds by a factor of 10 estimates of \cite{LehtoValtonen1996} and Dey et al.\ \cite{model2018}.   

\begin{figure}[ht]
\includegraphics[width=1\columnwidth]{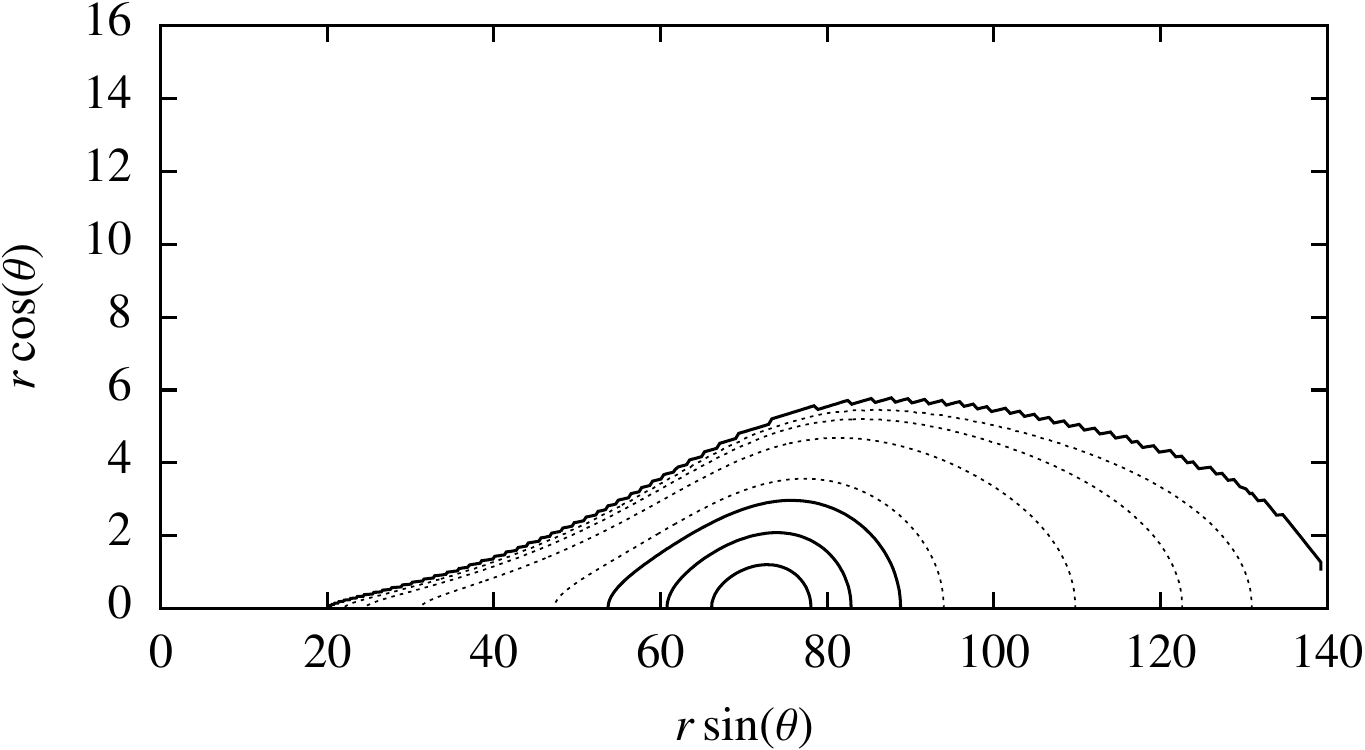}
\caption{\label{fig6} Here  $R_\mathrm{C1}=21.08 \, \mathrm{M}$, $R_\mathrm{C2}=141.03 \, \mathrm{M}$ and $p=K\rho^{4/3}$. Solid density isolines correspond to $\rho = (0, 2, 4, 6) \times 10^{-8}/\mathrm{M}^2$ (linear scale). Dotted density isolines correspond to $\rho = (10^{-11}, 10^{-10}, 10^{-9}, 10^{-8})/\mathrm{M}^2$ (logarithmic scale).}
\end{figure}  

In the next numerical solution, depicted in Fig.\ \ref{fig6}, we assume the circumferential radius of the innermost disk boundary $R_\mathrm{C1}=21.08 \, \mathrm{M}$ (or 3816 AU) and the circumferential radius of the outermost  boundary $R_\mathrm{C2}=141.03 \, \mathrm{M}$ (25535 AU). In standard SI units we have $R_\mathrm{C1}=5.71\times 10^{14} \, \mathrm{m} $ and  $R_\mathrm{C2}=3.82\times 10^{15} \, \mathrm{m}$. The mass of the torus is equal to $4.05\times 10^{-3} \, \mathrm{M}\approx 7.43\times 10^{7} \, \mathrm{M}_\odot$; this constitutes about 50\% of the mass of the secondary black hole. The height of the torus $h=5.63 \, \mathrm{M}$ (or  $1019 \, \mathrm{AU}  \approx 1.52\times 10^{14} \, \mathrm{m}$) exceeds several times the anticipated upper limit $1.4 \, \mathrm{M}$ \cite{LehtoValtonen1996,P2013, Pihajoki, model2018}.    

\begin{figure}[ht]
\includegraphics[width=1\columnwidth]{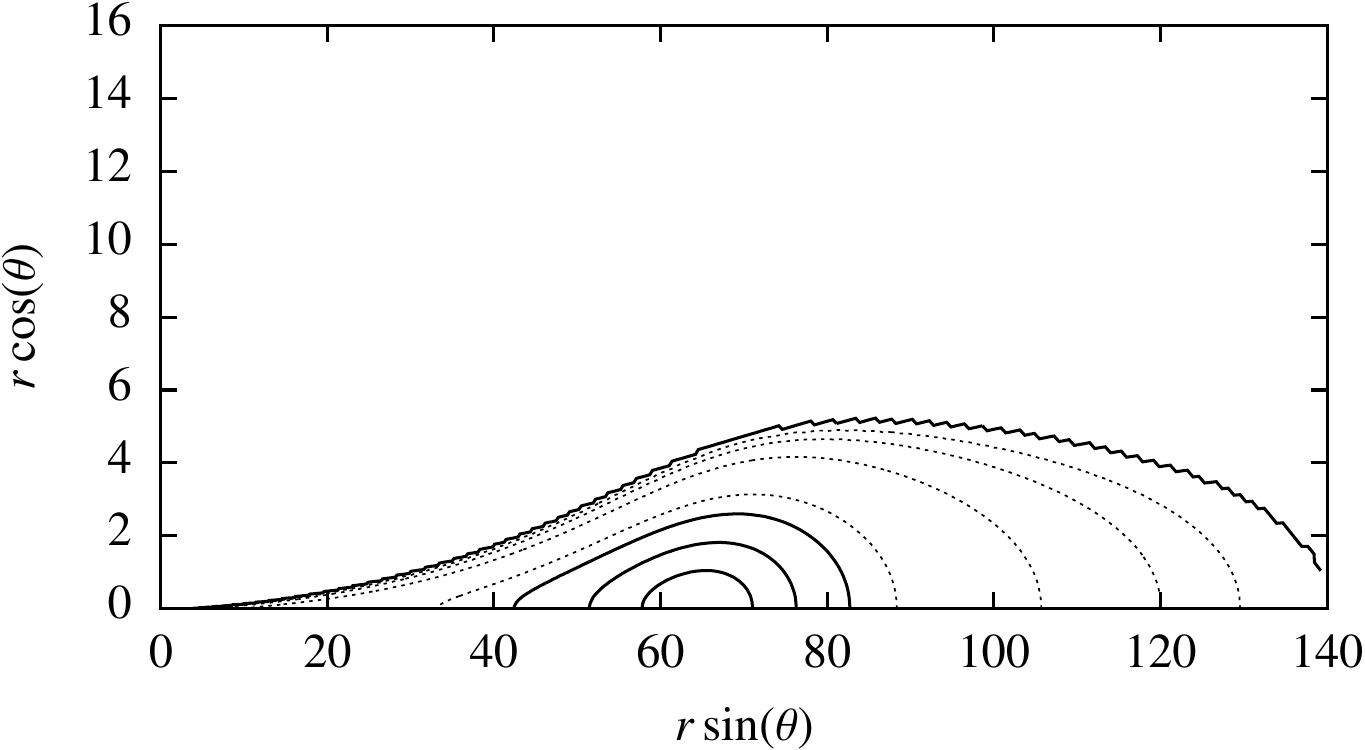}
\caption{\label{fig7} Here  $R_\mathrm{C1} =4.8 \, \mathrm{M}$, $R_\mathrm{C2} =141.06 \, \mathrm{M}$. The equation of state $p=K\rho^{4/3}$. Solid density isolines correspond to $\rho = (0, 2, 4, 6) \times 10^{-8}/\mathrm{M}^2$ (linear scale). Dotted density isolines correspond to $\rho = (10^{-11}, 10^{-10}, 10^{-9}, 10^{-8})/\mathrm{M}^2$ (logarithmic scale).}
\end{figure}  

The last of our numerical solutions is shown in Fig.~\ref{fig7}. The circumferential radii of the innermost and the outermost disk boundaries read  $R_\mathrm{C1} = 4.8 \, \mathrm{M}$ (or 870 AU) and $R_\mathrm{C2} =141.03 \, \mathrm{M}$ (25535 AU), respectively. In standard SI units we have $R_\mathrm{C1} =1.3\times 10^{14} \, \mathrm{m} $, $R_\mathrm{C2} =3.82\times 10^{15} \, \mathrm{m}$. The resulting mass of the toroid is equal to $3.52\times 10^{-3} \, \mathrm{M}$; that weight of more than $6.46 \times 10^7\, \mathrm{M}_\odot $ solar masses is almost half of the mass of the secondary black hole. Its height $5.09 \, \mathrm{M}$ (or c.\ 922 AU) exceeds almost four times expectations of \cite{LehtoValtonen1996} and Dey et al.\ \cite{model2018}.   

\section{Discussion}
\label{discussion}

In numerical solutions of Subsection \ref{numres1}, the toroids have geometric sizes --- horizontal and vertical --- close to those demanded by Valtonen and his coworkers. Their masses $M_\mathrm{d}$ are smaller by four orders than the mass $M$ of the primary black hole and by almost two orders from the secondary one, $M_\mathrm{sbh}/ M_\mathrm{d}\approx 50 \pm 10$. We should note that their maximum mass density is significantly smaller than required by Valtonen et al.\ for fluids satisfying the equation of state $ p=K\rho^{5/3}$. Numerical calculations show, that the equation of state $p=K\rho^{4/3}$ permits disks with smaller masses than in the case of the ``cold'' equation of state $p=K\rho^{5/3}$. The maximal mass density is of the same order as the mass density that was assumed in numerical simulations of \cite{Rezzolla} (see a discussion below). For that reason we regard the relativistic matter with $p=K\rho^{4/3}$ as a promising candidate for the future modelling of the disk in  OJ 287.

As pointed above, solutions of Subsection \ref{numres1} satisfy the required mass hierarchy $M\gg M_\mathrm{sbh}\gg M_\mathrm{d}$. This suggests that one can reduce the 3-body problem --- the black hole binary coupled to a torus --- to a chain of three simpler problems. It is reasonable to assume, taking into account the large value of the mass ratio $ M/M_\mathrm{sbh} \approx 100$, that the large-scale structure of a light rotating gaseous torus would result in interactions only with the heavy central black hole. The motion of the secondary black hole in turn would be dictated mainly by the primary Kerr-like black hole, since $M_\mathrm{sbh}\gg M_\mathrm{d}$. And finally, the mutual interaction between the disk and the secondary black hole would be important only locally during periodic encounters, when a general-relativistic variant of the Bondi-Hoyle-Lyttleton accretion model would be appropriate.

As argued by Valtonen and his coworkers, transits of the secondary black hole through the constructed torus would produce periodic flashes of radiation in OJ 287 with the luminosity of the order of $10^{13} \, \mathrm{L}_\mathrm{\odot}$. The mechanism of such a process has been described qualitatively in Newtonian hydrodynamics by Bondi, Hoyle and Lyttleton \cite{HL,BH} (see also a review of Edgar \cite{Edgar}). A quantitative numerical analysis of the passage of a black hole through a uniform fluid volume, within the framework of general-relativistic radiation hydrodynamics, has been done  by Zanotti et al.\ \cite{Rezzolla}. It is notable that this investigation, with a mass density similar to the maximal mass density in our examples described in Sec.\ \ref{numres11} and \ref{numres12}, gives the required luminosity of the order of $10^{13} \, \mathrm{L}_\mathrm{\odot}$ (cf.\ \cite{model2018}). The assumption of \cite{Rezzolla} --- that a tiny black hole traverses a large  uniform  gaseous medium --- is obviously not satisfied in our case. The mass density of our solutions is not constant, and the secondary black hole has a size close to 2 AU, while the maximal disk's thickness is smaller than 400 AU. The numerical analysis of the transition of the secondary black hole  through our disks would require suitable modifications, but it should yield the anticipated luminosity.  
 
In numerical examples described in Subsection \ref{numres2}, we fixed the same maximum mass density, as in \cite{LehtoValtonen1996, P2013} and \cite{model2018}. As it was already pointed out, in such a case the maximal disk's height has to become a part of the output data. It appears that the height of the tori obtained in this case is much higher --- even by a factor of 10 --- than assumed in the analysis of \cite{LehtoValtonen1996} and \cite{model2018}. Their profiles are shown in Figs.\ \ref{fig4}--\ref{fig7}. It is clear that these general relativistic tori are not thin, in contrast to the assumptions made by  Valtonen and his coworkers. That can mean that some characteristics obtained from these models  --- the luminosity or time scales of the interaction of the secondary black hole with disks --- would not agree with observations.

Notice also that the disks reported in Sec.\ \ref{numres2} are heavy --- their mass can exceed 3 to 4 times the mass of the conjectured second black hole, for the equation of state $p=K\rho^{5/3}$. The disks corresponding to $p=K\rho^{4/3}$ are less massive, but still their masses constitute a significant fraction of the mass of the secondary black hole. Thus the reduction of the 3-body problem to a set of two-body problems becomes problematic. In those circumstances the problem would have to be analysed in the full generality. One would pose initial data for the three bodies, possibly employing the obtained disk solution, and in principle track their evolution. In practise, this procedure would appear expensive numerically or even intractable.  
 
In conclusion, there are two arguments --- formal and factual (referring to observations) --- to rule out solutions with relatively dense disks interiors, $\rho_\mathrm{m} \ge 7.25\times 10^{-8}/\mathrm{M}^2$  ($\rho_\mathrm{m} \ge 1.33\times 10^{-7} \, \mathrm{kg/m^3}$) advocated in \cite{LehtoValtonen1996} and \cite{model2018}. 

\section{Summary}

The galactic nucleus in OJ 287 is interpreted as a binary black hole with a supermassive central black hole and a much lighter companion, and with an accretion disk surrounding the supermassive center. The geometrical sizes of the components and of their trajectories are estimated from observations filtered through a hybrid general-relativistic and Newtonian modelling \cite{BBHmodel1988,LehtoValtonen1996,Valtonen2007,Valtonen2008testGR,P2013,Pihajoki,model2018}.

We have shown, purely within the full general-relativistic setting, the existence of appropriate selfgravitating stationary tori that satisfy the mass hierarchy $M\gg M_\mathrm{sbh}\gg M_\mathrm{d}$. That this hierarchical ordering can be done, is the essential implicit assumption present in the former investigation \cite{BBHmodel1988,LehtoValtonen1996,Valtonen2007,Valtonen2008testGR,P2013,Pihajoki,model2018}, since it allows for the reduction of the description of a 3-body system to a triple of two-body systems. Thus our results support the validity of the aforementioned geometric picture of OJ 287.  

We did not investigate the production of the luminosity during transits of the secondary black hole through the disk. There exist appropriate models of the Bondi-Hoyle-Lyttleton accretion within general-relativistic radiation hydrodynamics \cite{Rezzolla}. We believe that their adaptation to gaseous interiors of our disk solutions --- or their modifications --- would yield luminosity characteristics that agree with observations.
 
\begin{acknowledgments}
This research was carried out with the supercomputer ``Deszno'' purchased thanks to the financial support of the European Regional Development Fund in the framework of the Polish Innovation Economy Operational Program (Contract no.\ POIG.\ 02.01.00-12-023/08). PM was partially supported by the Polish National Science Centre grant No.\ 2017/26/A/ST2/00530.
\end{acknowledgments}

\end{document}